\documentclass[twocolumn, aps, floatfix, superscriptaddress,showpacs]{revtex4-1}
\usepackage{dsfont}
\usepackage{amsmath}
\usepackage{enumerate}
\usepackage{amssymb}
\usepackage{graphicx}
\usepackage{wrapfig}
\usepackage{setspace}
\usepackage{cancel}
\usepackage{xcolor}

\def\XXint#1#2#3{{\setbox0=\hbox{$#1{#2#3}{\int}$}
     \vcenter{\hbox{$#2#3$}}\kern-.5\wd0}}

\renewcommand{\>}{\rangle}
\newcommand{\<}{\langle}

\begin{document}

\title{Coherence Factors Beyond the BCS Result}
\author{G. Gorohovsky}
\affiliation{Racah Institute of Physics, Hebrew University,
Jerusalem, Israel}
\author{E. Bettelheim}
\affiliation{Racah Institute of Physics, Hebrew University,
Jerusalem, Israel}

\date{\today}

\begin{abstract}
The dynamics of BCS (Bardeen-Cooper-Schrieffer) superconductors is fairly well understood due to the availability of a mean field solution for the pairing Hamiltonian, a solution which gives the quantum state of superconductor as a state of almost-free fermions interacting only with a condensate. As a result, transition probabilities may be computed, and expressed in terms of matrix elements of electron creation and annihilation operators
between approximate eigenstates. These matrix elements are also called 'coherence factors'. Mean-field theory is however not sufficient to describe all eigenstates of a superconductor, a deficiency which is hardly important in (or very close) to equilibrium, but one that becomes relevant  in certain out of equilibrium situations. We report here on a computation of    matrix elements (coherence factors) for the pairing Hamiltonian between any 'two-arc' eigenstates in the thermodynamic limit.  
\end{abstract}
\maketitle

\paragraph{Introduction}
A host of phenomena within the classical theory of Bardeen Cooper  and Schrieffer (BCS) of superconductivity may be explained  within the mean-field theory that BCS\ have put forth\cite{BCS1}. The success of BCS\ theory relies on the ability of mean-field theory to provide an approximation to a set of eigenstates and eigenvalues of a Hamiltonian that includes the pairing interaction responsible for superconductivity . Nevertheless, some out-of equilibrium phenomena require a fuller set of eigenstates than those that the mean field theory can provide.  Such out of equilibrium situations include quantum quenches\cite{Spivak:Levitov:Barankov,Kuznetsov2} (where the pairing interaction strength is suddenly changed making use, e.g.,  of the Feshbach resonance), but also problems related to non-equilibrium steady state\cite{Bettelheim:Multi:Gapped}. As an alternative approach to man-field, one may solve the pairing Hamiltonian, Eq. (\ref{Hamiltonian}),   exactly, as Richardson did many years ago \cite{Richardson}. We apply the solution here to large systems having in mind applications to non-equilibrium macroscopic superconductivity, probing a different set of problems than those  which are addressed by  a perhaps  more popular   application of Richardson's solution, namely the study of  small superconducting systems\cite{Braun:VonDelft,Imry:Schechter}.   

Although Richardson did not refer to the Bethe ansatz, his solution of the pairing Hamiltonian may be phrased in the language of the algebraic Bethe ansatz\cite{Amico:BCS:As:6:Vertex,vonDelft:Poghossian:Richardson:As:Bethe}. The problem of finding matrix elements of physical operators between exact eigenstates of a Hamiltonian within the Bethe ansatz approach is known to be a difficult problem. Nevertheless, recent progress in different models have been made \cite{Caux:Calabrese:Slavnov:1P:Dynamical:in:LiebLinger,Calabrese,Kitanine:XXZ:Form:Factors,Biegel:Karbach:Heisenberg,Kostov:Three:Point:N4SYM:Announce:PRL,Amico:Osterloh:Exact:Correlations,Gorohovsky:Bettelheim:Expectation:Values} based mainly on Slavnov's determinant representation of overlaps\cite{Slavnov}. 

 In this letter we report on  analytical expressions for properly coarse-grained matrix elements between exact eigenstates of the pairing Hamiltonian, Eq. (\ref{Hamiltonian}), in the thermodynamic limit. The derivation of the result will be published elsewhere \cite{Gorohovsky:Bettelheim:Preparation}, while here we only compare the result to numerics.  The matrix elements we report on can be used to study the dynamics of superconductors in far from equilibrium situations, in which the superconductor cannot be assumed to be in a state well described by the BCS\ eigenstates. The matrix elements we compute feature in the Fermi golden rule transition rates, which in turn find their way into a quantum Boltzmann equation formalism, and as such, our results may be applied for example to study, using a Boltzmann equation approach, the long time evolution after a quantum quench.

\paragraph{Model and Results.}
It is well established that a superconductor may be described by the following effective pairing Hamiltonian:

\begin{align} \label{Hamiltonian}
H = \sum_{  j , \sigma_j } \varepsilon_j c^\dagger_{j,\sigma_j} c_{j,\sigma_j} - G \sum_{ j,l} c^\dagger_{j,+} c^\dagger_{j,-} c_{l,+}c_{l,-}.
\end{align}
Here $(j,+)$ and $(j,-)$ denote the quantum numbers of time reversed pairs. For example, if $(j,+)$ denotes a state with wave number $\vec{k}$ and spin up, then $(j,-)$ denotes a state with wave number $-\vec{k}$ and spin down. We make the following assumptions with no loss of generality for the sake of simplicity. Namely, we assume that each level $j$ is only doubly degenerate, where $\sigma$ indexes the two degenerate states, $\sigma$ taking $+$ and $-$ as values. Furthermore, we assume uniform level spacing $\varepsilon_j  - \varepsilon_{j-1} = \iota$.

We will briefly review Richardson's solution and then give our expressions for matrix elements between exact eigenstates of the Hamiltonian. To describe an exact eigenstate, first note that only pairs of electrons are dynamical, while single electrons decouple from the dynamics. Indeed, suppose a single particle level at energy $\varepsilon_m$ is occupied with a single electron with either $+$ or $-$ spin, then the hamiltonian $H$ prescribes no dynamics for this electron. This means that a good quantum number, $M$,  is given by the number of single particle level which are singly occupied.  In addition to this, the non-dynamical nature of singly occupied levels, means that  for a given eigenstate $|v\>$, we may also associate a set of singly-occupied levels, $\varepsilon_{i_j}$, and the spin of the electron at the level, $\sigma_j$. 

Given the single-occupancy data, one must further classify all eigenstates with a given  single-particle occupation. This may be done by specifying a set of {\it rapidities}, namely, a set of $P$ complex numbers. We denote this set by $V$, where  $V=\left\{v_\mu\right\}_{\mu=1}^P$. The number $P$  is related to the total number of electrons, $N$,  in the system and $M,$ the number of singly occupied levels, as follows: $P=\frac{N-M}{2}$. The number $P$  may then be considered as the number of Cooper pairs in the system.   

\begin{figure}[h!!!]
\includegraphics[width=8cm]{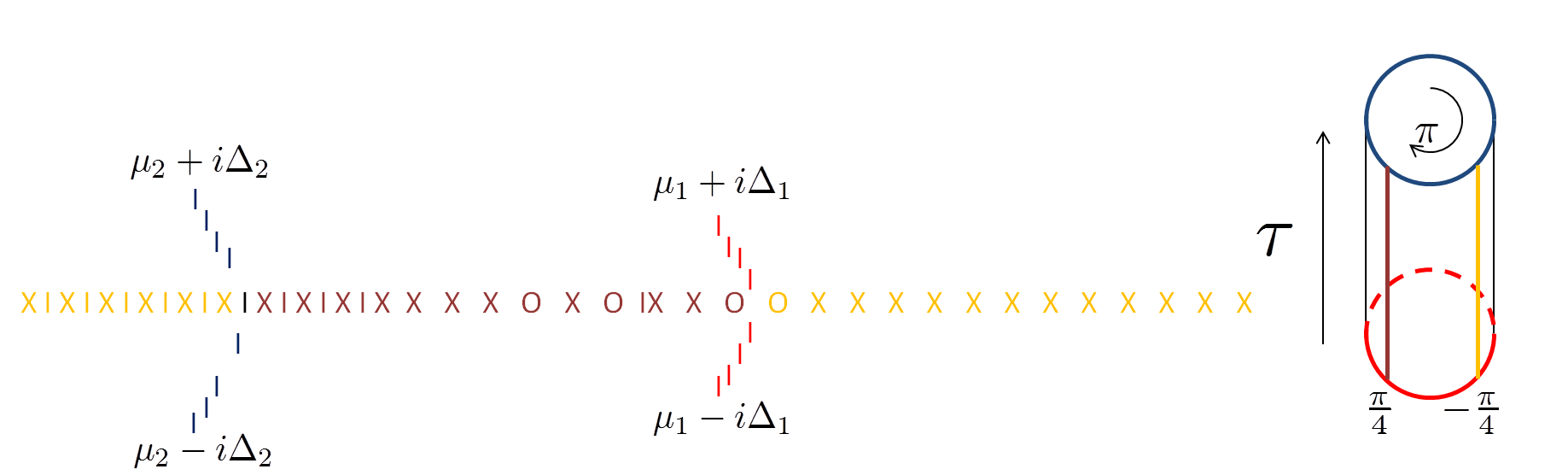}
\caption{A typical configuration of rapidities, the $I$ denote rapidities while the letter $X$ denotes unoccupied single particle levels the $O$'s denote singly occupied  levels. The mapping $\xi \to u_\xi$ maps the energy plane onto a cylinder of circumference $\pi$ and height $|\tau|$. The left arc is mapped to the base and the right arc to the top. The parts of the real axis outside and between the arcs are mapped onto the orange and brown lines, respectively.   \label{RichardsonConfigurationFigure}}
\end{figure}

The  rapidities  must satisfy Richardson's equations \cite{Richardson,Gaudin,Sierra}, namely for every $\mu, $  the following must be satisfied:
\begin{align}\label{Richardson}
\sideset{}{'}\sum_{\nu}\frac{2}{v_\mu-v_\nu} -\sum_i\frac{1}{v_\mu-\varepsilon_{i}}+\sum_j\frac{1}{v_\mu-\varepsilon_{i_j}}=\frac{2}{G}
\end{align}
The end result is that, for a given  single-electron occupancy and a set of rapidities, $V,$ one has an eigenstate of $H$ with energy 
$
E=\sum_\mu v_\mu+\sum_j \varepsilon_{i_j}.\label{spectrum}
$
We may think of the rapidities as living in complexified energy space. The single particle spectrum $\varepsilon_i$ constitutes a discrete set on the real axis in this space, while the $v'$s can in principle be found anywhere in this two-dimensional space. The Richardson equations, Eq. (\ref{Richardson}),  constraining the $v$'s lend themselves to a typical form of the solution. Namely, some of the $v$'s are found on the real axis in more-or-less arbitrary positions dispersed between  the single particle levels, while other $v$'s arrange themselves in arcs in the complex plane. There may be any number of arcs,  while in the present paper we restrict ourselves to the case where there is either $1$ or $2$ arcs. The number of arcs will be denoted by $k$, and the end points of the arcs by  $\{\mu_{i}\pm \Delta_{i}\}_{i=1}^k$. Fig. \ref{RichardsonConfigurationFigure} displays such a typical configuration with two arcs. 

In the thermodynamic limit we describe the distribution of single particle levels, singly occupied states and the rapidities by coarse-grained densities. We multiply all densities by the level spacing to obtain   'occupancy numbers'.  While the density of the $v$'s on the real axis is largely arbitrary, the density on the arcs may be found given the real-axis density of $v_\mu$'s and $\varepsilon_{i_j}$'s and the arc endpoints $\{\mu_{i}\pm \Delta_{i}\}_{i=1}^k$. We may then compute the matrix elements between two Richardson states each described by its real-axis density and its arc end-point. We denote such a state by $|n_+,n_-,n_V,\{\mu_{i}, \Delta_{i}\}_{i=1}^k\>.$ Here and below  $n_\alpha$, where $\alpha $ an take the 'values' $+, -$ or $V$ are  given in the following:
$n_\alpha(\varepsilon)= \frac{\iota}{\delta \varepsilon}\Xi_\alpha(\varepsilon) ,
$ 
where $\Xi_V(\varepsilon)$, $\Xi_{+}(\varepsilon)$ and $\Xi_{-}(\varepsilon)$ are respectively the number of rapidities, singly occupied levels with spin $+$ and singly occupied levels with spin $-$ in the segment $[\varepsilon-\frac{\delta \epsilon}{2},\varepsilon+\frac{\delta \epsilon}{2}]$ and $\delta \varepsilon$ is a coarse graining scale defined such that it is much larger than $\iota$, the level spacing, and much smaller than the scale at which densities change. We  define an excitation occupation number, $n(\varepsilon),$ as follows:
\begin{align}
n(\varepsilon)=n_+(\varepsilon) + n_-(\varepsilon) + 2n_V(\varepsilon).\label{nDef}
\end{align}
To see that $n$ is indeed an occupation number, note that, due to the relation between energy and the rapidities, $
E=\sum_\mu v_\mu+\sum_j \varepsilon_{i_j}.\label{spectrum}
$
Therefore, we may regard both the rapidities and the singly occupied levels of both spins as excitations. It is only the real rapidities which are counted in $n$, since only the real rapidities have an arbitrary density, and thus can be regarded as excitations. Since the number of rapidities corresponds to the number of Cooper pairs in the system the charge associated with them is double the charge of the singly occupied levels, and hence the factor $2$ in front of $n_V$ in (\ref{nDef}).

The state  $c^\dagger_{m\sigma}|n_+,n_-,n_V,\{\mu_{i}, \Delta_{i}\}_{i=1}^k\>$  will generally  have little overlap with states with significantly different occupation numbers, $n_\alpha$ and arc-endpoints. Thus denoting the in-state, $|\mbox{in}\> $, by $|n_+,n_-,n_V,\{\mu_{i}, \Delta_{i}\}_{i=1}^k\>,$ we compute the matrix element $\<\mbox{out}|c^\dagger_{m,\sigma}|\mbox{in}\>$, where the state  $|\mbox{out}\>$ may be considered to have the same density and arc-endpoints as  $|\mbox{in}\>$. Nevertheless, the object, $\<\mbox{out}|c^\dagger_{m,\sigma}|\mbox{in}\>,$ is not a diagonal matrix element since $|\mbox{out}\>$ may be different from the in-state on a microscopic scale. 

We describe the difference between $|\rm{in}\>$ and $\<\rm{out}|$ states  by two variables, $p$ and $l$. We claim that all order $1$ overlaps  are covered by the following values of $p$ and $l$.  The number  $p$ is any integer much smaller than $N$ counting how many more rapidities are on the left arc in the $|\mbox{out}\>$  state as compared to the $|\mbox{in}\>$ state. The number  $l$ is defined to be $1$ $(-1)$ if the $|\mbox{out}\>$ state has one excitation more (less)  next to $\varepsilon_m$ as compared to $|\mbox{in}\>$. Note that $c^\dagger$ ostensibly creates an excitation, so  naively $l=1$, however, a well known feature of superconductivity is that a condensation of a pair may accompany the creation of an excitation.   The latter corresponds here to a rapidity leaving the vicinity of $\varepsilon_m$ and joining an arc -- a process which brings down $l$ to $-1$. 

\begin{figure}[h!!!]
\includegraphics[height=5cm]{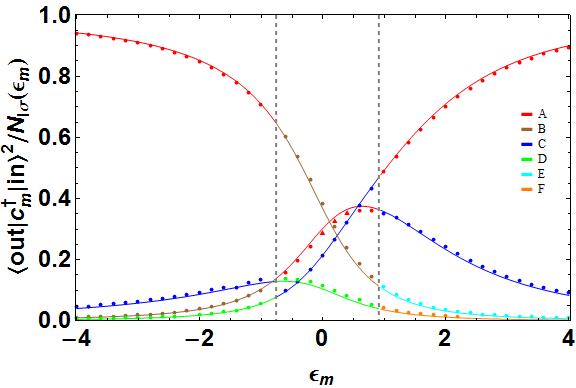}
\caption{
\label{ResultGraph}
 Matrix element square between an in- and out-state divided by  $N_{l,\sigma}$. Colors correspond to a out-states according to:
{\bf A}: $l=1$ $p=0$;
{\bf B}: $l=-1,$ $p=1$;
{\bf C}: $l=-1,$ $p=0$;
{\bf D}: $l=1,$ $p=-1$;
{\bf E}: $l=1$, $ p=1$; {\bf F}: $l=-1$, $p=-1$.
 Arc endpoints are given by: $\Delta_1=2.2534$, $\mu_1=0.8257,$ $\Delta_2=1.0612$ and $\mu_2=-0.0499$.  The numerics were done by situating $P$  rapidities solving Richardson's equations on  two arcs. In addition single particle levels have been evenly distributed at the intervals $[2,4]$, $[-4,-2]$  and $[-0.5,0.5]$ with level spacing denoted by $\iota $. Circles:  $P=385$, $\iota=\frac{1}{348}$. Triangles: $P=577$, $\iota=\frac{1}{522}.$    Richardson's equations were solved numerically to obtain $\mu_i$ and $\Delta_i$  with  $P=641$,  $\iota=\frac{1}{580}$.}
\end{figure}

To write the main result of this paper, we define $N_{l,\sigma}=\delta_{l,1}-l\left(n_{l\sigma}+n_V\right)$, which allows us to write:   
\begin{align}\label{IntroductionResults}
&\<\mbox{in};l,p|c^\dagger_{m\sigma}|\mbox{in}\>^2=\frac{\pi^2 l N_{l,\sigma}\sin^{-2}\left[u_{\varepsilon_m}+l(p\tau-u_\infty)\right]  }{2\omega^2 R_4({\varepsilon_m})}
\end{align}
where:\begin{align}
&R_{4}(\xi) =\prod_{j=1}^2 \sqrt{ (\xi - \mu_j)^2 +\Delta_j^2} &\omega =\int_{\mu_1-i \Delta_1}^{\mu_1+i \Delta_1} \frac{1}{R_{4}(\xi')} d\xi \nonumber\\
&u_\xi =\frac{\pi}{ 2\omega }\int_{\mu_1+i \Delta_1}^\xi \frac{1}{R_{4}(\xi')} d\xi & \frac{\tau}{2}=u_{\mu_2+i\Delta_2}\quad u_\infty=\lim_{\xi\to\infty}u_\xi.  \nonumber\end{align}
All expressions must be taken by drawing branch cuts for the function $R_4(\varepsilon)$ such as to coincide with the arcs (the  shape of the arcs is given by the solution to the Richardson equations, which are tractable in the thermodynamic limit\cite{Sierra,Gaudin}). The path of integration  in the definition  of $u_\xi$ should not cross the branch cuts, but may wind around it. The path of integration in the definition of $\omega$  is drawn slightly to the right of the branch cut. A graphical depiction of the our main result, Eq. (\ref{IntroductionResults}), is shown in  Fig. \ref{ResultGraph}.

Note that the mapping $\xi\to u_\xi$ is ambiguous because the path of integration in the definition of $u_\xi$ may wind around the arcs. As a result, $u_\xi$ is defined modulo addition of $m  \pi$, where $m$ is an integer.   We denote the Abelian group of integer translations by $\pi$ as $\mathbb{Z} \pi$ and the set of points on the branch cuts as $\mathcal{C}$.  Due to the ambiguity,   the mapping $\xi \to u_\xi$, must be understood as  taking $\mathbb{C}\setminus\mathcal{C}$ to $\mathbb{C}/\mathbb{Z}\pi$. The function $\sin^{-2}$ is well defined on $\mathbb{C}/\mathbb{Z}\pi$ due to its $\pi$ periodicity. The quotient $\mathbb{C}/\mathbb{Z}\pi$ has the topology of an infinite  cylinder. The image of the mapping is actually a finite portion of this cylinder extending over $0<\mbox{Im}(u_\xi)<|\tau| $. The portions of the real axis between and outside the  arcs are mapped onto $\mbox{Re}(u_\xi)=\pm\frac{\pi}{4} $ , respectively, see Fig. \ref{Richardson}.   

It is important to note that result, Eq.  (\ref{IntroductionResults}), is to be understood as being coarse grained in the following sense. We average the left hand side of (\ref{IntroductionResults}) over all in and out states with the same coarse grained occupation numbers $N_{l,\sigma}$.

\begin{figure}
\includegraphics[width=5cm]{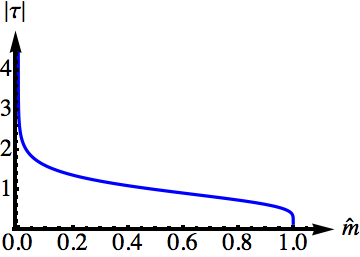}
\caption{The ratio $|\tau|$ as a function of the modulus $\hat m$. \label{DrawModulus}}
\end{figure}

\paragraph{Properties of the result.}We note now a few properties of the solution. The  matrix element square decays as $e^{-\left|p\tau\right|}$ as $p\to\pm\infty$. The ratio  $\tau$ becomes larger as the distance between the arcs becomes larger as compare to the size of the arcs.  In fact $\tau$ is a function, $\tau(\hat m)$,  of the cross-ratio of the end-points, $\hat m=\frac{(\Delta_1+\Delta_2)^2-(\Delta_1-\Delta_2)^2}{(\Delta_1+\Delta_2)^2+(\mu_1-\mu_2)^2}<1 $. The variable $\hat m$ called the elliptic modulus and ranges from $0$ when either one of the arcs vanishes ($\Delta_i=0$)  or when the distance between the arc tends to infinity  ($|\mu_1-\mu_2|\gg\Delta_i$) to $1$ when both arcs coincide ($\Delta_1=\Delta_2$ and $\mu_1=\mu_2$) . The ratio $\tau$ goes to zero as $\hat m$ approaches zero and diverges as $\hat m$ approaches $1$. The cross ratio is invariant to translations and dilations of the set of arc end points. Fig. (\ref{DrawModulus}) gives a plot of $\left|\tau\right|$ as a function of $\hat m$.   The explicit functional dependence may be written through elliptic integrals, $
\tau(\hat m)= i \frac{K(1-\hat m)}{K(\hat m)},
$ where $K$ is the complete elliptic integral of the first kind. 

As the arcs draw far apart from each other, or when one of the arcs becomes very small, the problem is effectively that of one-arc, namely the BCS expressions should hold. In this limit $\hat m$ goes to $1$,    the imaginary number $\tau$ goes to infinity, and the result indeed converges to the usual BCS expression. To see this, first consider that all $p\neq0$ are suppressed by factors of $e^{-|p\tau|}$. We need then only to consider the case $p=0$. Next assume without loss of generality that  $\varepsilon_m$ close to $\mu_1.$ In the limit $\tau\to\infty$  we have for $\xi$ close to the first arc, the estimate:
\begin{align}
&i \Delta_1e^{-i 2 u_{\xi}} \simeq\mu_1-\mu_2+\frac{R_2(\mu_2)\left[R_2(\mu_2)-R_2(\xi)\right]}{\mu_2-\xi} \end{align}
where $R_2(\xi) = \sqrt{(\xi-\mu_1)^2+\Delta_1^2}$.  Additionally, making use of the estimate $\omega\simeq \frac{\pi i}{R_2(\mu_2)}$, we  obtain:
\begin{align}
\<\mbox{in};l,0|c^\dagger_{\varepsilon,\sigma} | \mbox{in}\>^2=N_{l,\sigma}\left[1-l\frac{\varepsilon-\mu_1}{\sqrt{(\varepsilon-\mu_1)^2+\Delta_1^2}}\right]\nonumber.
\end{align}
The latter expression is the BCS result. 

We denote four points on the real axis $\varepsilon_{A^\pm}$, $\varepsilon_{B^\pm}$, where $A$ and $B$ denote the points the right and the left arcs meet the real axis, respectively, and the $\pm$ superscript denote approaching those points from the left and the right respectively. We have the following relation for matrix elements evaluated at these points:
 \begin{align}
&N^{-1}_{l,\sigma} \<\mbox{in};l,p|c^\dagger_{A^\pm,\sigma} | \mbox{in}\>=N^{-1}_{-l,\sigma}\<\mbox{in};-l,p|c^\dagger_{A^\mp,\sigma} | \mbox{in}\> ,\label{JoinUP}\\&N^{-1}_{l,\sigma} \<\mbox{in};l,p|c^\dagger_{B^{\pm},\sigma} | \mbox{in}\> =N^{-1}_{-l,\sigma}\<\mbox{in};-l,p+l|c^\dagger_{B^{\mp},\sigma} | \mbox{in}\>.\nonumber
\label{JoinUP}
\end{align}
The equality follows formally from  (\ref{IntroductionResults}), but may also be intuitively understood by noting the fact that the out-state described on the left and right hand side in Eqs. (\ref{JoinUP}) are in fact indistinguishable.  

\paragraph{Comparison with numerics.} In Fig. \ref{ResultGraph} the matrix element square is drawn for a few values of $l$ and $p$. One can see curves corresponding to different $p$ and $l$ joining up at the points at which the arcs cross the real axis, corresponding to relations (\ref{JoinUP}).

Fig. \ref{ResultGraph} depicts also a comparison with numerical results, where, first, Richardson's equations (\ref{Richardson}) were solved numerically,  secondly, the matrix elements were computed using Slanvnov's formula, which requires computing a large determinant\cite{Zhou:Links:McKenzie:Gould,Calabrese,Calabrese}.       

\paragraph{Relation to expectation values. } Eq. (\ref{IntroductionResults}) allows to compute expectation values. By inserting a complete set of states one can compute the expectation value of $\hat{N}_m\equiv\sum_\sigma  c^\dagger_{m,\sigma} c_{m,\sigma}$ . The result is as follows:
\begin{align}
&\<\hat N_m\> =\sum_{l,\sigma}
 \frac{\pi^2lN_{l,\sigma}(\varepsilon_m)}{2\omega^2R_4({\varepsilon_m})} 
 \sum_{n}
\sin^{-2}
\left[ u_{\varepsilon_m}+l(p\tau-u_\infty)\right].\nonumber
\end{align}
The sum over $p$  may be taken explicitly by considering the right hand side as a function in the complex $u_{\varepsilon_m}$ plane. For a given value of $n$ the sum on the right hand side has double poles at $-l(p\tau-u_\infty)+\pi j$ for any integer $j$. Summing over $p$ we get a function which has double poles  at a two dimensional lattice of points, $-lu_\infty+\mathds{\pi Z}+  \tau \mathds{Z}$. This largely fixes the sum to be given by the Weierstrass $\wp$-function. In fact applying the standard lore of elliptic functions yields the following formula for $\<N_m\>$:
\begin{align}
\<\hat N_m\!\!-\!\!1\>\! =\frac{\wp(u_\infty-u_{\varepsilon_m})+\wp(u_\infty+u_{\varepsilon_m})+2\frac{\eta}{\omega}}{2R_4({\varepsilon_m})}(n(\varepsilon_m)\!\!-\!\!1),\nonumber
\end{align}
where $\wp$ is the Weierstrass $\wp$-function with period $\pi$ and $\tau$, while the constant $\eta$ is a standard notation  in the theory of Weierstrass elliptic functions \cite{abramowitz:stegun}.

The result for $\<\hat{N}_m-1\> $ agrees with the result obtained in Ref. [\onlinecite{Gorohovsky:Bettelheim:Expectation:Values}]. More generally, expectation values are given in terms of elliptic functions (doubly periodic functions in the complex $u_\varepsilon$ plane),  a  conclusion reached either by invoking semi-classics [\onlinecite{kuznetsov},\onlinecite{Kuznetsov2}] or through an exact approach\cite{Gorohovsky:Bettelheim:Expectation:Values}. All such expectation values may be alternatively written as sums over matrix elements of single fermionic operators, by insertion of the resolution of the identity. These matrix elements are found here to be given by trigonometric functions. Thus,  in general, the expansion of expectation values in terms of matrix elements is an expansion of elliptic functions in terms of trigonometric functions. 

\paragraph{Summary.} In conclusion we note, that, in principle, one can find the matrix elements of bilinear operators in fermions by a semi-classical method. Namely, the time dependence of the such bilinear operators may be found in the semi-classical limit invoking the methods developed in  \cite{Volkov1974,Spivak:Levitov:Barankov,Kuznetsov2,kuznetsov} . After these are obtained, one may Fourier transform the time-dependent expressions to obtain matrix elements. The method does not lend itself, however, to the computation of the matrix elements of single fermionic operators, such as the ones reported on here, which prompts us to resort to an approach based on an exact diagonalization of $H$. 

The results we presented here are applicable to the treatment of the dynamics of out-of-equilibrium superconductors. The importance of coherence factors in the theory of non-equilibrium BCS superconductivity is well recognized. Here we find such coherence factors for a more broader set of eigenstates, including such eigenstates which have special relevance to non-equilibrium superconductors\cite{Bettelheim:Multi:Gapped}. 

The methods used to compute matrix elements, which will be detailed in a forthcoming publication\cite{Gorohovsky:Bettelheim:Preparation}, bear importance more generally as an example of an analytical  computation of matrix elements in systems with a Bethe-ansatz solution in the thermodynamic limit. Such computation are rare, but recent progress has seen some success in finding such expectation values in different contexts\cite{Calabrese,Caux:Calabrese:Slavnov:1P:Dynamical:in:LiebLinger}. Here we         provide  completely analytical computations (as opposed to a semi-numerical approach \cite{Calabrese,Amico:Osterloh:Exact:Correlations}) in a  model which contains a macroscopic string of rapidities (the arcs). Such models were discussed in the condensed matter theory context by Sutherland \cite{Sutherland:MacroString}, and have received renewed interest within the context of integrability in the AdS/CFT correspondence\cite{Sever:Gromov:Vieira:Integrable:FieldTheories,Ksotov:Three:Point:N4SYM:Details,Kostov:Three:Point:N4SYM:Announce:PRL}.  
\paragraph{Acknowledgement}
We acknowledge discussion with P. Wiegmann and B. Spivak. 
EB is grateful for the hospitality at the University of Cologne, where the work has been completed. This work has been supported by the Israel Science Foundation (Grant No. 852/11) and by the Binational Science Foundation (Grant No. 2010345).

\bibliographystyle{apsrev4-1}
\bibliography{mybib}
\end{document}